\documentclass[usenatbib]{mnras}
\usepackage[utf8x]{inputenc}
\usepackage[english]{babel}
\usepackage[T1]{fontenc}
\usepackage{array}
\usepackage{amsmath}
\usepackage{graphicx}
\graphicspath{{EPS/}}
\usepackage{float}
\usepackage{fixltx2e}
\usepackage{xcolor}
\usepackage{hyperref}

\usepackage{tikz,xcolor,hyperref}

\definecolor{lime}{HTML}{A6CE39}
\DeclareRobustCommand{\orcidicon}{%
	\begin{tikzpicture}
	\draw[lime, fill=lime] (0,0) 
	circle [radius=0.16] 
	node[white] {{\fontfamily{qag}\selectfont \tiny ID}};
	\draw[white, fill=white] (-0.0625,0.095) 
	circle [radius=0.007];
	\end{tikzpicture}
	\hspace{-2mm}
}

\foreach \x in {A, ..., Z}{%
	\expandafter\xdef\csname orcid\x\endcsname{\noexpand\href{https://orcid.org/\csname orcidauthor\x\endcsname}{\noexpand\orcidicon}}
}

\hypersetup{pdfstartview=FitH, colorlinks=true, linkcolor=blue, filecolor=blue, urlcolor=blue}

\newcounter{ourcount}
\title[Acceleration of protons in the Fermi bubbles]{On the high-energy protons regular acceleration in the Fermi Bubbles}
\author[Istomin Ya.~N., Gunya A.~A.]{Istomin Ya.~N. $^{1}$\thanks{E-mail: \href{mailto:istomin@lpi.ru}{istomin@lpi.ru}}{\orcidB{}}, Gunya A.~A. $^{1}$\thanks{E-mail: \href{mailto:aagunya@lebedev.ru}{aagunya@lebedev.ru}}{\orcidA{}} \\
	${1}$ P.N.~Lebedev Physical Institute, Leninsky Prospect 53, Moscow 119991, Russia \thanks{Website: \href{https://www.lebedev.ru/ru/}{www.lebedev.ru}} \\}

\begin{document}
	\date{}
	\pagerange{\pageref{firstpage}--\pageref{lastpage}}
	\pubyear{2021}
	\maketitle
	\label{firstpage}
	
\begin{abstract}
	
    We propose the model of the global structure of the electromagnetic fields in the Fermi bubbles (FBs), which makes possible the proton regular acceleration up to ultra-high energies. The poloidal and the toroidal magnetic fields, as well as the radial electric field, turn to have a structure similar to fields that exist in jets ejected out from active galactic nuclei (AGN). A powerful source of relativistic particles observed in the centre of the Galaxy and associated with the rotating supermassive black hole (SMBH) Sgr A* can energize the FB and keeps its active for a long time. The absence of accretion onto a BH and thus the absence of a relativistic jet does not mean that there is no loss of rotational energy of BH. In the case of FB, the energy lost by BH can keep the FB activity. The regular FBs structure could be formed by inheritance from a relativistic jet that presumably existed in the active past of the Galaxy $ 10^7 $ years ago, or by processes near the Galaxy centre existing during the entire life cycle of the Galaxy. The acceleration of protons in electromagnetic fields of FB are found up to energies $ E_{max} \simeq 10^{17} $ eV, which explains the observed radiation of FB in the gamma range, as well as the emission of high-energy neutrinos.

\end{abstract}


\section{Introduction}

    The Fermi bubbles (FBs) are sphere-like formations above and below of the Galaxy plane, was first discovered in 2010 \citep{2010ApJ...724.1044S}. They have the radius of $ R \simeq 3 $ kpc, comparable with the radial dimension of the central region of the Galaxy. This structures apparently filling with a rarefied relativistic plasma which have intensive and sharp radiation spectrum in the MeV gamma range. 
    
    There are several scenarios of the FBs origin. One of the first intriguing assumption of the FBs origin was the scenario of evolution of the relativistic jet remnant. Presumably, relativistic jets existed during the recent activities of the Sgr A* about $ 10^7 $ years ago. This assumption originating many interesting discussions about FBs inheritance from relativistic jets \citep{2020ApJ...894..117Z}, \citep{2012ApJ...761..185Y}, \citep{2012ApJ...756..182G}. The FBs can presumably fill with matter from jets originating by the SMBH vicinity. The expansion process could occur at the front of a shock wave, which collided with the external environment and formed the boundaries of the currently observed FBs. On the other hand, in the paper \citep{2012ApJ...753...61S} discussed by Su \& Finkbeiner, the evidence was given of the possibility of a gamma jet existence, which could also lead to the filling of such a structure as FBs by hot particles. There is another possibility of the FBs formation due to the tidal disruption of stars \citep{2017EPJWC.14504004C}, \citep{2014NuPhS.256..179C}, \citep{2011ApJ...731L..17C} and the acceleration of individual particles by the arising shock waves. The FBs primary structure origin by supernova explosions in the central molecular zone (CMZ) also is possible \citep{2014MNRAS.444L..39L}. The young stars formation and the stellar winds can also affect on to the FBs origin, including into the CMZ life cycle \citep{2019MNRAS.490.4401A}. The paper \citep{2011ApJ...731L..17C} also described a scenario of FBs periodic feeding by supernova explosions. Various interesting scenarios for the FBs formation were also discussed by Mertsch \& Petrosian in the paper \citep{2019A&A...622A.203M}. Comparative analysis of the possible FBs origin, in general, is of great interest for explaining its nature since discovery in 2010. It's also interesting to note that all scenarios which mentioned beyond can be complement each other. FBs can be really formed from relativistic jets, but intensive periodic (on average, once every $ 10^4 $ years) processes with supernovas can be really energizing FBs. Stellar plasma injected by supernova shock waves can be one of the important additional channels for the energy supply of a stationary bubble formed from a jet.
    
    In this paper, the authors basically admit that in the case of the scenario of the FBs inheritance from jets, the expansion of the jet remnants for a time of $ 10^7 $ years could preserve the similarity of the currents that existed earlier in the jets and, therefore, the similarity of the magnetic field topology. Of course, it's not excluded, that another scenario can also be responsible for FBs origin.
    
    As a result, as in the case of the jet inheritance scenario or in the case of primary formation by supernovae, the electromagnetic fields formed to present time inside the FBs can lead to particle acceleration by regular mechanism, which described in the present work and by the stochastic mechanism \citep{2015ApJ...799..112C}, \citep{2020ApJ...904...46K},\citep{2012ApJ...746..116C}, \citep{2014ApJ...790...23C}, \citep{2015ApJ...804..135C}.
    
    In the case of considering the FBs as a stationary or quasistationary structure with a regular electromagnetic field, a particle acceleration mechanism is possible, which produces  particles of high energies of the order of $ 10^{17} $ eV by a global electric field. It is the acceleration of protons rather than electrons that is considered in this work, since they experience almost no synchrotron radiation. Moreover, among high-energy protons in FBs can exist particles preliminary accelerated in the magnetosphere of Sgr A* \citep{2020MNRAS.492.4884I} to energies of $ 10 ^ {15} $ eV, which are practically not deflected by the magnetic fields. 
 
    The possible structure of regular electromagnetic fields in Fermi bubble is described in the section (\ref{section2}). The equations of motion of protons and their solution are presented in the section (\ref{section3}). A discussion of the results and their interpretation in the context of the physical parameters of the FBs is presented in the section (\ref{section4}).  
	
\section{Fields structure}\label{section2}
	
    Before considering the structure of the FBs electromagnetic field, it is necessary to make several remarks about assumptions of the evolution and stationarity of the FBs form, since they key to define the model described in the work.
    
    Since the scenario of the formation of FBs from a jet used in this work, the issue of the FBs supposed evolution should be briefly considered. After slowing down the intensive accretion, the jets matter may have begun to diffusely expand in the Galaxy halo. Moreover, due to the exponential decrease of the halo magnetic field with a height above the plane of the Galaxy disk $ z $, $ B \propto \exp (-z / z_0) $, where the characteristic size $ z_0$ is $\simeq 2$ kpc, the initial expansion occurs generally in the vertical direction due to the increase in the diffusion coefficient $ D $ together with a decrease in the magnetic field. Effectively, the vertical expansion is convective rather than diffusional, with a velocity $ u = dD / dz $. At the base, $ z \simeq 0 $, the vertical velocity is of $ u \simeq 10^3 $ km/s \citep{2013Natur.493...66C}. The velocity $ u $ is large enough, so that the energy of the vertical motion $ u^2 / 2 $ exceeds the gravitational potential $ g_d $ of the Galaxy stellar disk with mass $ M_d \simeq 6 \cdot 10^{10} M_\odot $ and the radius $ R_d \simeq 15 $ kpc. For $ z <R_d $ in the axial region, the value of $ g_d $ is $\simeq 2G M_d z / R_d^2 \simeq 5 \cdot 10 ^ {13} (z / z_0)$ $cm^2 / s^2 $, where $ G $ is the gravitational constant. Although velocity $ u $ decreases to the value of, $ u \simeq 10^2 $ km/s, at an altitude above $ z_0 $ the influence of gravitational potential remains insignificant. In the transverse direction, perpendicular to $ z $, the movement is diffusional. Therefore, the present bubbles have a shape elongated in the vertical direction. A detailed calculation of the FBs shape evolution was made in the paper \citep{2014NewA...27...13I}.

    The general background of the FBs is considered as quasistationary with very slow diffusional expansion. All perturbations against such a background are considered to be small and do not significantly determine the global structure of the field. There are several general considerations indicating that if turbulence does arise, then it is not significant in the meaning that the amplitudes of the disturbances are small compared to stationary values. They are: (1) the fact that bubbles are formed in the process of gas diffusion, and diffusion is a dissipative process, the perturbations should decrease exponentially with the decrement $ \gamma_d \simeq 4 \pi^2 D / l^2 $, where $ l $ is the size of the perturbations. In any case, practically all perturbations, $ l <2 \pi (Dt)^{1/2} \simeq 2 \cdot 10^{22} $ cm, will have time to slowing down to present time; (2) as we propose, the structure of electromagnetic fields in bubbles is similar to the structure of the jets fields, and as we know, jets are stable; (3) the instability of a structure means its disequilibrium. The growth of perturbations should lead either to the formation of a new equilibrium structure with the absence of strong perturbations, or to a strongly turbulent state (like Kolmogorov turbulence). In the first case, the level of turbulence is small, and the equilibrium configuration of  fields should be close to the structure proposed in our work based on general physical considerations. In the second case of strong field turbulence, the FBs shape would not have regular structure, which does not seen to be observed.
    
    Several more assumptions were also made: (1) since disturbances or changes in the FBs structure are not considered, the electromagnetic field is stationary, the electric field is potential; (2) the FBs periphery, that is FBs boundaries,  is not considered in the work, since they do not make significant changes in the proton acceleration process.
    
    Due to assumptions of the symmetry both of the FBs in the sections ({\ref{section2}}) and ({\ref{section3}}) only one Fermi bubble (FB) is considered.
    
	The FB structure can be considered axisymmetric with a good approximation. This means that all quantities in FB are independent of the azimuthal angle $ \phi $, the angle around the axis of the Galaxy. Therefore, the magnetic field in FB can be represented as
	
\begin{equation}\label{Bfields}
	B_\rho = - \frac{1}{\rho} \frac{\partial f}{\partial z}, \, B_z = \frac{1}{\rho} \frac{\partial f}{\partial \rho}, \, B_\phi = \frac{1}{\rho} g. 
\end{equation}

    Here the quantity $ f(\rho, z) $ is the flux of the poloidal magnetic field depending on the cylindrical coordinates, the radius $ \rho $, and the vertical distance $ z $. The relation $ f (\rho, z) = const $ describes the magnetic surfaces on which lines of the magnetic field of the poloidal magnetic field lie. The toroidal magnetic field $ B_\phi (\rho, z) = g (\rho, z) / \rho $ is created by the  poloidal electric currents. In the case when these currents lie on magnetic surfaces, $ g (\rho, z) = g (f) $.

    We will assume that the environment filling the FB is ideal, i.e. the electric field $ {\bf E} $ in the Galaxy frame satisfies the relation $ {\bf E} = - [{\bf u} \times {\bf B}] / c $. Here the value of $ {\bf u} $ is the velocity of the matter. At times significantly shorter than the lifetime of FB, $ \simeq 10^7 \, $ years, the motion of the medium is a rotation with the angular velocity $ \Omega $. As a result,
\begin{equation}\label{erho}
	E_\rho=-\frac{\Omega}{c}\frac{\partial f}{\partial \rho}, \, E_z=-\frac{\Omega}{c}\frac{\partial f}{\partial z}, \, E_\phi=0.
\end{equation}
    Since the electric field is potential, $ {\bf E} = - \nabla \Psi $, then the angular velocity $ \Omega $ is only a function of the magnetic surface, $ \Omega = \Omega (f) $. The potential of the electric field is
    $$
	\Psi=\frac{1}{c}\int^f \Omega(f')df'.
	$$
    The electric current $j_\phi $  in the azimuthal direction consists of the motion of electrons and protons, $ j_\phi = en (u_{\phi i} -u_{\phi e}) $. The quantity $ n $ is the density of the quasineutral plasma. The electrons presented in the FB, regardless of the central engine, are undoubtedly magnetized, i.e. their cyclotron radius is much smaller than the size of the FB. Therefore, they rotate in the azimuthal direction with the angular velocity $ \Omega (f) $, $ u _ {\phi e} = \rho \Omega (f) $. As for the protons, they can be on the verge of magnetization. Protons were supplied or are delivered to the FB through the central paraxial region $ \rho \simeq 0 $. In general, their azimuthal velocity can be found from the angular momentum conservation law,
    $$
	\rho u_{\phi i}+\frac{e}{c m_p}f=const=L.
	$$
    The magnetic field flux is zero at $ \rho = 0 $, therefore $ L = 0 $, and
    $$
	u_{\phi i}=-\frac{e}{c m_p}\frac{f}{\rho}.
	$$
	Thus, the toroidal electric current is	
\begin{equation}
	j_\phi=-en\left(\frac{e}{c m_p}\frac{f}{\rho}+\rho\Omega(f)\right).
\end{equation}
    Plasma is frozen into the poloidal magnetic field $ B_p = [(\nabla f) ^ 2]^{1/2} / \rho $. Therefore, the plasma density $ n $ is proportional to $ B_p $,
	$$
	n=\frac{n_0}{\rho B_0}\left[\left(\frac{\partial f}{\partial \rho}\right)^2+\left(\frac{\partial f}{\partial z}\right)^2\right]^{1/2}.
	$$
    Here the values of the plasma density $ n_0 $ and the magnetic field strength $ B_0 $, respectively, are their mean values in the FB. Maxwell's equation gives the equation for the magnetic field flux $ f $
\begin{eqnarray}
	&&\rho\frac{\partial}{\partial\rho}\left(\frac{1}{\rho}\frac{\partial f}{\partial\rho}\right)+\frac{\partial^2 f}{\partial z^2}= \\
	&&=\frac{4\pi n_0 e}{cB_0}\left(\frac{e}{c m_p}\frac{f}{\rho}+\rho\Omega(f)\right)\left[\left(\frac{\partial f}{\partial \rho}\right)^2+\left(\frac{\partial f}{\partial z}\right)^2\right]^{1/2}. \nonumber
\end{eqnarray}
    The FB shape is simplified now as a cylinder of radius $ R $, independent of the height $ z $. We introduce the dimensionless quantities (prime),
	$$ 
	f=B_0R^2 f', \, \rho=R \rho', \, z=R z', \, \Omega=c\Omega'/R,
	$$
	$$
	{\bf B}=B_0{\bf B}', \, {\bf E}=B_0{\bf E}'.
	$$
    Omitting the primes, we obtain a dimensionless equation, which determines the dimensionless flux $ f $
\begin{equation}\label{flux}
	{\Tilde{\Delta}} f=\left(\frac{\omega_p R}{c}\right)^2\left(\frac{f}{\rho}+\frac{c\rho\Omega(f)}{R\omega_c}\right)\left[\left(\frac{\partial f}{\partial \rho}\right)^2+\left(\frac{\partial f}{\partial z}\right)^2\right]^{1/2} \nonumber
\end{equation}
\begin{equation}	
    {\Tilde{\Delta}}=\rho\frac{\partial}{\partial\rho}\left(\rho^{-1}\frac{\partial}{\partial\rho}\right)+\frac{\partial ^2}{\partial z^2}. 
\end{equation}

    The equation (\ref{flux}) includes the values that characterize the FB environment. These are the ion plasma frequency, $ \omega_p ^ 2 = 4 \pi n_0 e ^ 2 / m_p $, and the ion cyclotron frequency, $ \omega_c = e B_0 / c m_p $. Since the $ \omega_p^2 \simeq 2 \cdot 10^6 \, n \, s^{- 2} $, the factor on the right-hand side of the equation (\ref{flux}) is large, $ (\omega_p R / c)^2 >> 1 $. On the other hand, $ \omega_c \simeq 10^{- 1} \, s^{- 1} $. Therefore, with good accuracy, the equation (\ref{flux}) is greatly simplified,
\begin{equation}
	f=-\frac{c}{R\omega_c}\rho^2\Omega(f).
\end{equation}
    This ratio seems to mean the absence of a toroidal electric current, and thus a poloidal magnetic field. However, the situation is similar to the phenomenon of quasineutrality, when the absence of charge density in a plasma does not mean the absence of an electric field. It should be noted here that for $ \Omega> 0 $, the flux of the poloidal magnetic field is negative, $ f <0 $, i.e. the vertical magnetic field $ B_z $ in the central region is negative, while for $ \Omega <0 $, the flux is positive, $ f> 0 $. 
    
    The angular plasma rotation frequency $ \Omega $, which is constant in the axial region, must vanish at the boundary of FB, $ \rho = 1 $. Therefore, let us choose the dependence of the rotation frequency on the radius $ \rho $ in the form $ \Omega = a (1- \rho)^2 $, where $ a $ is a positive constant, $ a> 0 $. The presence of differential rotation $ \Omega = a (1- \rho)^2 $ is necessary for global electric field. In this model assumed that FB fields directly connect with SMBH magnetosphere fields. The SMBH Sgr A* has definitely low, but not zero specific dimensionless angular momentum $ j \simeq 0.1 $ \citep{2020ApJ...901L..32F}. As for the FB, in the work \citep{2013Natur.493...66C}, the interpretation of experimental data in the context of the rotation of the FB base was also discussed. The dependence in the form $ (1- \rho) ^ 2 $, not $ (1- \rho) $, is connected with the fact that not only the radial electric field $ E_\rho $, $ E_\rho \propto \Omega $ , but also its derivative regarding the radius, must vanish at the boundary of FB.
    
    That is, not only the electric field, but also the electric charge density $ \rho_e $ at the boundary is equal to zero. As a result, the dependence of the poloidal magnetic field $ f $ flux on the radius $ \rho $ has the form
$$
	f=-a\frac{c}{R\omega_c}\rho^2(1-\rho)^2.  
$$
	
\begin{figure}
    \centering
    \includegraphics[width=80mm]{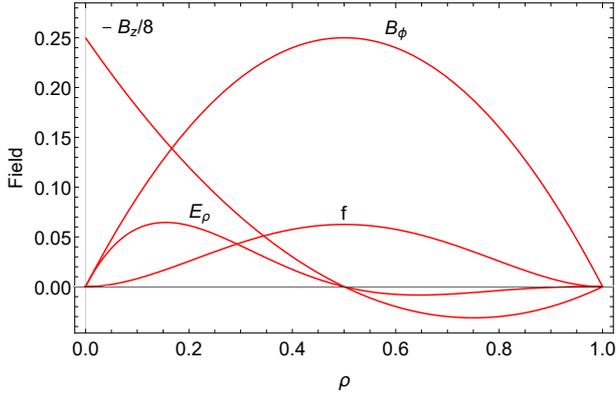}
    \caption{Dependences of electromagnetic fields and poloidal magnetic flux on the radial coordinate $ \rho $. The vertical magnetic field, $B_z \propto -(1-\rho)(1-2\rho)$; the toroidal magnetic field, $B_\phi\propto \rho(1-\rho)$; the radial electric field, $E_\rho \propto \rho(1-\rho)^3(1-2\rho)$; the poloidal magnetic flux, $f\propto\rho^2(1-\rho)^2$. The influence of fields at the periphery of the FB is neglected.}
    \label{1-2}
\end{figure}
    The $ f (\rho) $ dependency is shown in the figure (\ref{1-2}). The flux vanishes at the radius of FB. This means that the longitudinal, along the axis of rotation, magnetic field $ B_z $ changes its sign: in the inner region, $ \rho <1/2 $, the field is directed in one direction, in the outer one - in the opposite direction. The total flow through FB is zero. This is very different from the magnetic field in the magnetosphere of a SMBH, where the flux in a separate hemisphere is not zero, but has different signs in different hemispheres (split-monopole). Thus, the longitudinal magnetic field $ B_z = (\partial f / \partial \rho) / \rho $ is
\begin{equation}\label{bz}
    B_z=-2a\frac{c}{R\omega_c}\left(1-\rho\right)\left(1-2\rho\right).
\end{equation}

    An addition explanation should be made here. Microwave images (see, for example, Fig. 3 in Carretti \cite{2013Natur.493...66C}), clearly show the boundary between the FB and the surrounding media of the Galactic halo. There is no radiation outside the FB and the background is black. This means that there is no synchrotron and inverse Compton radiation from accelerated electrons in the halo, with the exception of individual sources. There are either no accelerated electrons or a sufficiently strong magnetic field. The absence of a magnetic field leads to the conclusion that there is no magnetic field in the medium surrounding the FB, as well as at the boundary between the FB and the halo matter. The probability that the magnetic field is significant and there are no accelerated particles is small due to the fact that accelerated electrons themselves are accelerated by turbulent electric and magnetic fields, both in the medium and in regular electric and regular and random magnetic fields of shock waves. Due to the continuity of the electromagnetic field at the boundary between the FB and the Halo conducting medium, the electromagnetic field at the FB boundary $\rho=1$ is equal to zero. Here we can say that free charges and currents in the halo do not allow the FB fields to penetrate into interstellar space, just as a field do not penetrate into an ideal superconductor. The use of MHD calculations in FB modelling, assuming that FBs are created by matter ejections from the central BH or its vicinity (Yang \cite{2012ApJ...761..185Y}; Barkov \cite{2014A&A...565A..65B}), does not lead to the conclusion that there are no electromagnetic fields at the FB boundary. This is due to the strong non-stationarity of the process. In the case of the FB, as the remnant of a jet from a black hole, the expansion of the FB proceeds slowly diffusion (see \cite{2014NewA...27...13I}). In addition, here the question arises what is meant the FB border? Let us remind one more time that in the article we consider the regular structure of the FB, and thus the regular electromagnetic fields inside. If the region near the boundary is turbulent and radiation is observed from it, this does not mean that there is a regular magnetic field at the boundary.

    It should be noted here that the dependence $ \Omega (f) $ has a different form in the regions $ \rho <1/2 $ and $ \rho> 1/2 $, but they are switched together at $ \rho = 1/2, \, y = 1 $
    $$
	\Omega = a\left[1+\left(1-y\right)^{1/2}\right]^2/4, \,
	$$
	$$
	y=4\left(-\frac{f R\omega_c}{a c}\right)^{1/2}, \, y<1, \, \rho<\frac{1}{2};
	$$
	$$
	\Omega = a\left[1-\left(1-y\right)^{1/2}\right]^2/4, \,
	$$
	$$
	y=4\left(-\frac{f R\omega_c}{a c}\right)^{1/2}, \, y<1, \, \rho>\frac{1}{2}.
	$$
    The toroidal magnetic field $ B_\phi $, on the one hand, is proportional to the radius $ \rho $ at small radii, and on the other hand, it should vanish at $ \rho = 1 $ at the boundary of FB. The derivative of $B_\phi$ regarding the radius $ \rho $ at the boundary must also be zero, which means that there is no longitudinal electric current density $ j_z $ at the boundary. So,
\begin{equation}\label{bphi}
	B_\phi=\alpha\rho(1-\rho)^2,
\end{equation}
    where the value of $ \alpha $ is a constant, which, as we will see below, is proportional to the total longitudinal electric current. The density of the longitudinal electric current $ j_z $ is equal to $ j_z = [\nabla \times B_\phi] c / 4\pi$,
$$
    j_z=\frac{1}{2\pi}c\alpha\left(1-\rho\right)\left(1-2\rho\right).
$$
    It is interesting to note that the longitudinal electric current $ j_z $ has the same dependence on $ \rho $ as the vertical magnetic field $ B_z $ (\ref{bz}). The current vanishes at $ \rho = 1/2 $, and flows in different directions at $ \rho <1/2 $ and $ \rho> 1/2 $. Its structure resembles that of an electric current in a jet injected out of AGN. This indirectly confirms our hypothesis that FB is inherited from the jet that was previously outflowing from the centre. FB is not just a formation once inflated of a matter of a jet or a matter from destroyed stars. The total longitudinal electric current $ I $, flowing in one direction, and then closing at the front boundary of FB, is equal to
\begin{equation}\label{cur}
    I=2\pi\int_0^{1/2}j_z\rho d\rho=\frac{\alpha}{32}c B_0 R.
\end{equation}

    The electric field has only a radial component $E_\rho$ (\ref{erho}), ($E_\phi = E_z =0$),
$$
    E_\rho=2a^2\left(\frac{c}{R\omega_c}\right)\rho\left(1-2\rho\right)\left(1-\rho\right)^3.
$$
	Let's introduce the dimensionless parameter of the electric field $\beta=2(ac/R\omega_c)^2(R\omega_c/c)$, then
\begin{equation}
	E_\rho=\beta\rho\left(1-2\rho\right)\left(1-\rho\right)^3.
\end{equation}
	The total potential difference $ U $ between the axis and the boundary of FB is
\begin{equation}\label{ubeta}
	U=\frac{\beta}{60}B_0 R.
\end{equation}
	Thus, the electromagnetic fields in FB are
\begin{eqnarray}\label{fields}
	&& B_z =-\delta (1 - \rho)(1-2\rho), \nonumber \\
	&& B_\phi = \alpha \rho(1 - \rho)^2, \\
	&& E_\rho = \beta \rho(1 - \rho)^3 (1-2\rho). \nonumber
\end{eqnarray}

    The constant $\delta$ is the amplitude of the vertical magnetic field $B_z$, $\delta = 2ac/R\omega_c$ (\ref{bz}). Using the definition of $a$ through $\beta$, $a=(\beta R \omega_c /2c )^{1/2}$, we have $\delta=(2\beta c/R\omega_c)^{1/2}$. Thus,  electromagnetic fields in FB (\ref{fields}) are defined by two constants $\alpha$ and $\beta$, which are proportional to the electric current $I$ (\ref{cur}) and the voltage $U$ (\ref{ubeta}) correspondingly. For the moderate values of $\beta$ the amplitude of $B_z$  is small, $\delta<<1$, and the magnetic field is mainly toroidal. It seems due to diffusive expansion of the jet remnant up to the large distance $R$.

    Note that the negative sign in the expression (\ref{fields}) for the vertical magnetic field $ B_z $ takes place in the case of positive rotation $ \Omega> 0 $. For $ \Omega <0 $, the sign is positive. In what follows, without loss of generality, we will assume $ \Omega> 0 $.
		
\section{Particle acceleration}\label{section3}

    The maximum energy that a charged particle can reach during regular acceleration by an electric field in the magnetosphere of an SMBH depends on the intensity of a poloidal magnetic field near a horizon. Here, the rotating SMBH plays the role of the unipolar inductor \citep{1977MNRAS.179..433B}. For Galaxy, the voltage generated by a Sgr A* is of the order of $ U \simeq 10^{15} $ V. However, a particle in the magnetosphere of SMBH in inactive phase does not pass the full potential difference $ U $, and the maximum Lorentz factor of protons is $\gamma_{max} = (eU / m_p c ^ 2) ^ {1/2} \simeq 4 \cdot 10^5 $ \citep{2020MNRAS.492.4884I}, i.e. their energy is of $ 4 \cdot 10^{14} $ eV. Here, the quantities $ e $ and $ m_p $ are the charge and mass of the proton, respectively. Total proton energy $ E = eU $ achieved in the FB, where the total SMBH potential difference ends. This energy corresponds to $ E_{max} \simeq 10^{17} $ eV. The same scenario authors consider for the AGN \citep{PhysRevD.102.043010}, where proton gains energy in the SMBH magnetosphere up to the value of $ m_p c ^ 2 (eU / m_p c^2)^{2/3}$ \citep{2020MNRAS.492.4884I}. The rest of the energy $ eU $, proton gains already in the region of the jet itself. 
    
    In the FB the motion of a proton in an electromagnetic field is described by the equations
\begin{eqnarray}\label{lorenz}
    &&\frac{d{\bf p}}{dt}=e\left({\bf E}+\frac{1}{c}\left[{\bf v B}\right]\right),  \nonumber \\ 
    &&\frac{d{\bf r }}{dt}=\frac{{\bf p}}{m_p\gamma}, \\ 
    &&\gamma^2=1+\frac{p^2}{m_p^2 c^2}. \nonumber
\end{eqnarray}
    Here $ \textbf{r} $ and $ {\bf p} $ are coordinates and momentum of a charged particle, $ \gamma $ is its Lorentz factor. We introduce dimensionless coordinates and time, as well as dimensionless momentum   
\begin{eqnarray}\label{dimensionless}
    && t=\frac{R}{c} t', \,  \rho=R \rho', \, z=R z', \nonumber \\
    && {\bf p}=\frac{R\omega_c}{c}m_p c p', \gamma = \frac{R\omega_c}{c} \gamma'.
\end{eqnarray}
    The ratio $ c / \omega_c $ is the cyclotron radius of a non-relativistic particle. It is much smaller than the radius of FB $ R, \, c / \omega_c R_B << 1 $. Omitting the prime indices, we pass to the equations of particle motion in the fields $ B_z, \, B_\phi, \, E_\rho $ (\ref{fields})
\begin{eqnarray}\label{move}
    &&\frac{dp_\rho}{dt} = \frac{p_\phi^2}{\rho \gamma} - \delta \frac{p_\phi}{\gamma}(1 - \rho)(1-2\rho) - \nonumber \\
    && - \alpha \rho(1 - \rho)^2 \frac{p_z}{\gamma} + \beta \rho(1 - \rho)^3 (1-2\rho), \nonumber \\
    &&\frac{dp_\phi}{dt}=-\frac{p_\rho p_\phi}{\rho \gamma} +\delta (1 - \rho)(1-2\rho) \frac{p_\rho}{\gamma},  \nonumber \\
    &&\frac{dp_z}{dt}=\alpha \rho(1 - \rho)^2 \frac{p_\rho}{\gamma},  \\
    &&\frac{d\rho}{dt}=\frac{p_\rho}{\gamma}, \nonumber \\
    &&\frac{dz}{dt}=\frac{p_z}{\gamma}.  \nonumber		
\end{eqnarray}

    The first terms on the right-hand sides of the first and second equations are inertial forces. The system of equations of motion (\ref{move}) has integrals of motion: energies $ {\cal E} = const $ and angular momentum $ {\cal L} = const $,
\begin{eqnarray}\label{conserve} 
    &&{\cal E} = \gamma - \beta \Psi(\rho), \nonumber \\
    &&{\cal L} = \rho p_\phi + \frac{\delta}{2}\rho^2(1 - \rho)^2.
\end{eqnarray}
    Here 
$$    
    \Psi(\rho) = \frac{1}{3}(1-\rho)^4\left(\rho^2-\frac{1}{5}\rho-\frac{1}{20}\right)+\frac{1}{60}=
$$
$$
    \rho^2\left(\frac{1}{3}\rho^4-\frac{7}{5}\rho^3+\frac{9}{4}\rho^2-\frac{5}{3}\rho+\frac{1}{2}\right)
$$
    is the potential of the radial electric field.
   
    We are interested in the acceleration of protons in the FB to energies much higher than the energy of the particles initially located in the formed bubbles or penetrating into the FB from the Galaxy disk. As it can be seen from the ratios (\ref{dimensionless}), the maximum hypothetical acceleration in FB is equal to $\gamma'\simeq 1, \,\gamma\simeq R\omega_c/c\simeq 3\cdot 10^{10}$.
    
    FB rotates, that experimentally determined \citep{2013Natur.493...66C}, with differential angular velocity $\Omega = a(1 - \rho)^2$. A rotating magnetic field creates a radial electric field $E_\rho$, which for magnetized particles leads to their corotation with FB. For particles of sufficiently high energy  (a large cyclotron radius), there is also a motion along the electric field, i.e. their acceleration. That is the mechanism of a regular acceleration in an electromagnetic field.

    From the conservation of the angular momentum (\ref{conserve}) for particles accelerated from the initial values of $ \rho \simeq 0 $, we have
$$
    p_\phi=-\frac{\delta}{2}\rho(1 - \rho)^2.
$$
    From the third equation of the system (\ref{move}) we obtain
$$
    p_z=\frac{1}{2}\alpha\rho^2\left(1 - \frac{4}{3}\rho + \frac{1}{2}\rho^2\right).
$$ 
    From the ratio $\gamma^2=p^2+(c/\omega_c R)^2$, $c/\omega_c R<<1$, knowing the components of the momentum $ p_\phi $ and $ p_z $, we find the radial momentum of a particle    
\begin{eqnarray}
    p_\rho^2=\beta^2 \Psi^2 - p_\phi^2 - p_z^2 = \nonumber \\
    \frac{1}{4}\beta^2\rho^4\left(\frac{2}{3}\rho^4-\frac{14}{5}\rho^3+\frac{9}{2}\rho^2-\frac{10}{3}\rho +1 \right)^2- \\
    \frac{1}{4}\alpha^2\rho^4\left(1-\frac{4}{3}\rho+\frac{1}{2}\rho^2\right)^2-\frac{1}{4}\delta^2\rho^2\left(1-\rho\right)^4. \nonumber
\end{eqnarray}
    It can be seen that the acceleration of particles in FB is possible only under the condition $ | \beta |> | \alpha | $, i.e. provided that the amplitude of the electric field is greater than the amplitude of the magnetic field. Otherwise, the particle injected from the paraxial region is captured by the magnetic field and does not propagate into the region of sufficiently large radii, where the potential $ \Psi (\rho) $ reaches large values. In addition, the acceleration of particles starting from small values of the radius $ \rho_0 $ (axial region) is possible at sufficiently large values of $ \beta> \rho_0^{- 1} $.
\begin{figure}
    \centering
    \includegraphics[width=0.9\columnwidth]{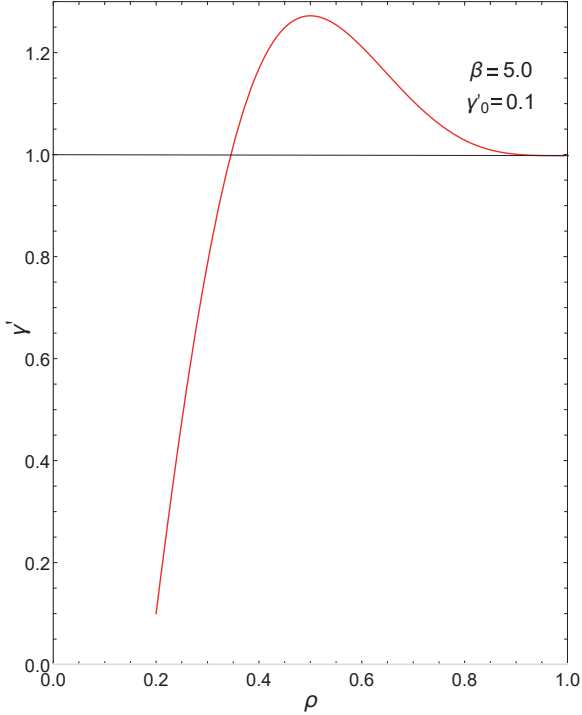}
    \caption{Dependence of the Lorentz factor $ \gamma' $ on the radial coordinate $ \rho $. The value of $ \gamma' $ is taken as unity for a particle that has passed the entire FB radius, $\gamma' = \beta\Psi(\rho=1)=\beta/60$}
    \label{1-3}
\end{figure}
    From the figure (\ref{1-3}) it can be seen that the maximum energy of protons is reached not at the boundary of FB, but at its middle at $ \rho = 1/2 $, and the value of the energy in the middle is $ \simeq 1.3 $ times energy value at the boundary of FB. This is since the radial electric field $ E_\rho $ at $ \rho> 1/2 $ becomes negative (see Fig. \ref{1-2}). The value of the electric potential $ U $ traversed by a particle injected from the paraxial region $ \rho \simeq 0 $ into FB to its boundary $ \rho = 1 $ is (see the formula \ref{ubeta}), $ U = \beta B_0 R / $ 60. For simplicity, we considered FB to be a cylinder with constant radius $ R \simeq 10^{22} $ cm. However, this maximum size is only achieved at a sufficiently high altitudes $ z \simeq R $. The size $ R_0 $ of the FB, based on the Galaxy disk, is less than $ R $, $ R_0 \simeq 2 \cdot 10^{21} $ cm. Since the boundary of FB supposed is equipotential, otherwise a strong electric current would flow along the boundary, we should put $ U = \beta B_0 R_0 / 60 \simeq 1.7 \cdot 10 ^ {14} V (B_0 \sim 5 \cdot 10 ^ {- 6} G) $. This value is almost equal to the potential difference generated by the rotating SMBH of the Galaxy, $ \simeq 2 \cdot 10 ^ {14} $ V \citep{2020MNRAS.492.4884I}, which corresponds to the maximum proton energy  $\simeq 10^{17} $ eV.  
    
\begin{figure}
    \centering
    \includegraphics[width=1.0\columnwidth]{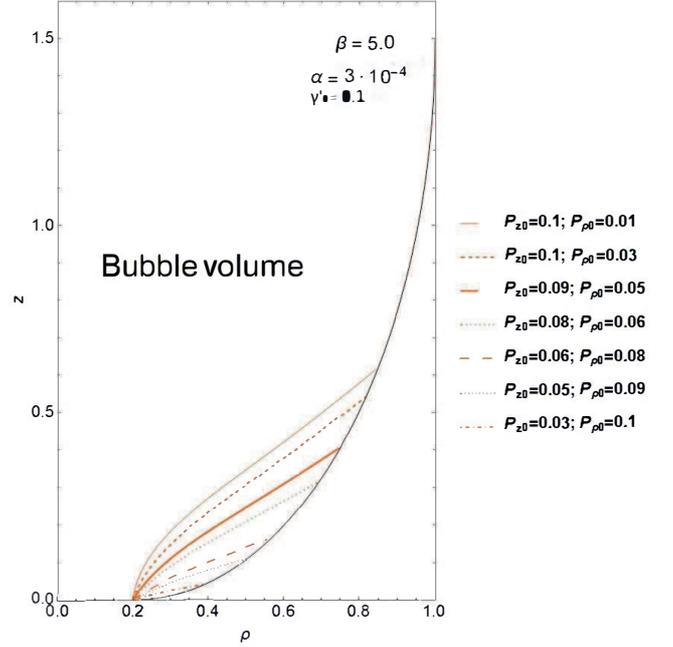}
    \caption{Trajectories of accelerated protons depending on their initial momenta $ p_{z0} $, $ p_{\rho 0} $. The thin line marks the boundary of the FB.}
    \label{1-4}
\end{figure}

    Let us now determine the parameters of the electric field $ \beta $ and electric current $ \alpha $ in FB. The radial electric field $ E_\rho $ (\ref{fields}) and the toroidal magnetic field $ B_\phi $ (\ref{fields}) create a flux of electromagnetic energy along the $ z $ axis
\begin{equation}
    s_z=\frac{c}{4\pi} E_\rho B_\phi=\frac{\alpha\beta c B_0^2}{4\pi}\rho^2(1-\rho)^5(1-2\rho).
\end{equation}
    The flux  inside FB is directed in the positive direction in the paraxial region $ \rho <1/2 $, and outward at the periphery, $ \rho> 1/2 $. Energy is flowing into the FB, reflected from the upper boundary, and spreads throughout the FB region. The total flux of energy entering into FB is equal to
\begin{equation}
    S_z=\frac{\alpha\beta c B_0^2 R^2}{2}\int_0^{1/2}\rho^3(1-\rho)^5(1-2\rho)d\rho.
\end{equation}
    Integration gives   
$$
    S_z=2.46\cdot 10^{-4} \alpha\beta B_0^2 R^2 c.
$$
    The approximate energy flux $ J $ emitted by FB is equal to $ J \simeq 10 ^ {38} $ erg/s. It mainly consists of TeV gamma photons ($ \simeq 4 \cdot 10 ^ {37} $ erg/s) and X-ray photons (1-10 keV \citep{2012MNRAS.423.3512C}), whose luminosity is an order of magnitude higher \citep{2011PhRvL.106j1102C}. Using the standard values of the magnetic field and radius of FB, $ B_0 \simeq 10 ^ {- 5} $ G and $ R \simeq 10^{22} $ cm, we get $\alpha\beta\sim 1.35\cdot 10^{-3}$, that actually determine dimensionless power.

\section{Discussions}\label{section4}

    This model depicted that FBs really can have a suitable environment for efficient acceleration of protons to the high energy. The main population of protons preliminary accelerated in the magnetosphere Sgr A* and further accelerated to high energy is located in the regions of the FBs closest to the disk plane. But the largest energy in FB is achieved in the central region of FB.

    This model depicted that FBs really can have a suitable environment for efficient acceleration of protons to the high energy. The main area of the population of protons accelerating to high energies is located at the base of the FBs. The structure of electromagnetic fields of FBs can really have similar structure to the structure of relativistic jets fields. In this case, Sgr A* can have direct relation to FBs. The absence of intense accretion of matter onto the SMBH, and thus the absence of a jet injected from its magnetosphere, does not mean that there is no loss of the rotation energy of the SMBH. 
    
    Observations show that flux $ \simeq 10^{38} $ erg/s is emitted from the GC in the form of relativistic protons with energies up to $ \simeq 10 ^ {15} $ eV \citep{2016Natur.531..476H}. This energy is sufficient to fully or partially feed the FBs. The resulting value of the maximum energy $ \gamma_{max} $ achieved by protons in the FBs, $ \gamma_{max} \simeq R_0 / \omega_c / 60 c \simeq 10^8 $ ($\simeq 10^{17}$ eV), corresponds to the crossing of the particle through the entire potential difference between the FBs axis and its boundary. At an arbitrary radius $ \rho ( \rho <1 )$, the energy is less and equals, $\gamma'= \beta \Psi (\rho) $.
    
    The  accelerated particles outgoing from the initial coordinate $ \rho_0 $, located on the base of the FBs, $ z \simeq 0 $, and propagating up to the point $ \rho $, form the entire energy spectrum. Suppose that the density of primary particles $ n (z = 0) $ at the base of the FBs is homogeneous, $ n (\rho) = const(\rho) = n $. Then the number of particles $ N $ spreading from values $ \rho = \rho_0 $ to the final values $ \rho $ is 
$$
    dN=n\rho_0 d\rho_0.
$$
    The relationship between the initial value of the coordinate $ \rho_0 $ and the current value $ \rho $ is given by the trajectory
$$
    \frac{dz}{d\rho}=\frac{p_z}{p_\rho}.
$$ 
    For a fixed height level $ z $, we thus have a connection between $ d \rho $ and $ d \rho_0 $,
$$
    d\rho_0=\frac{p_\rho(\rho_0) p_z(\rho)}{p_\rho(\rho)p_z(\rho_0)}d\rho.
$$
    On the other hand, $ d \gamma = \beta (d \Psi / d \rho) d \rho $. Therefore, we finally have
$$
    \frac{d N}{d\gamma}=\frac{n\rho_0}{\alpha}\frac{p_z(\rho}{p_\rho(\rho)}\frac{d\rho}{d\Psi}, \gamma=\beta\Psi(\rho).
$$
    The value of $\rho_0$, generally speaking, is of the order of the radius of the base of FBs onto the Galactic stellar disk $R_0$.
    Here it is considered that for axial region, small $ \rho_0 $, $ p_\rho (\rho_0) / p_z (\rho_0) =  \beta / \alpha $.
    From the expressions of $ \Psi (\rho) $ for small values of the radius, $  \rho <1/2 (\Psi \propto \rho^2) $ and for values closed to the maximum $ \rho \simeq 1  ( \Psi \propto (1- \rho)^4) $, it is seen that the spectrum of accelerated particles at $ \gamma <\gamma_{max} / 4 $ is hard, 
$$
    \frac{dN}{d\gamma}\propto \gamma^{-1},
$$
    while at higher energies, $ \gamma_{max} / 4 < \gamma <\gamma_{max} $ becomes softer,
$$
    \frac{dN}{d\gamma}\propto \gamma^{-7/4}.
$$
    It should be noted that this index is close to the index -2, which is universal for acceleration by strong shock waves.
    
    At the time, we have weak observation data of the high-energy protons from FBs. There is an objective reason for this. The high-energy proton concentration $ n_L $ near the light cylinder of the SMBH Sgr A* magnetosphere can be estimated from the consideration of the superiority of the energy of the electromagnetic field over the energy of accelerated protons, hence the density of protons is $ n_L = B_L ^ 2/4 \pi \gamma_{max} mc ^ 2 $. The magnetic field $ B_L $ for the light cylinder Sgr A* is of the order of 1 G, the maximum Lorentz factor of the high-energy protons is $ \gamma_{max} \simeq 10 ^ 5 $. From where, $ n_L \simeq 10 ^ {-3} $ cm$^{-3}$. For the base of the FBs, where high-energy protons are accelerated according to the trajectories in the figure (\ref{1-4}), the concentration of such protons will already be of the order of $10^{- 17}$ cm$^{-3}$. However, their total number $N$ in the entire volume of FB, $\simeq 10^{67}$ cm$^{3}$, $N\simeq 10^{50}$ is pretty big. Of course, their density is significantly lower than the average FBs plasma density at the base $n_H \simeq 10^{-3}$ cm$^{-3}$. This imposes significant restrictions on the observation of such particles. Direct registration of high-energy protons accelerated in FB is also difficult because their trajectory makes a significant angle to the plane of the Galaxy in which the observer is located. The next interesting sign may be the background itself. Photohadronic scenario obviously is not acceptable due to small photon number. On the contrary, proton-proton scenario partially can explain gamma background. But this requires: (1) average time for collision significantly more than FBs lifetime in the jet remnant scenario $t_{pp} \simeq 1/c n_H \sigma_{pp} \simeq 10^8$ yr, where the cross-section is, $\sigma_{pp} \simeq 30$ mb; (2) high injection rate of protons $L_p \simeq N E_{max}/t_{pp} > 10^{38}$ erg/s. 
    
    On the existence of high-energy protons in FBs can also indicate the registration of energetic neutrinos. At the time, it's the most objective and acceptable sign of the high-energy protons acceleration. Some events registered by IceCube \citep{2020PhRvD.101l3023B}, \citep{2018Galax...6...47R}, \citep{2016PhRvD..93a3009A}, \citep{2014PhRvD..90b3016L} were associated with high statistical accuracy with FBs, which corresponds to the area of their occurrence. In particular, the regions of neutrino registration at latitudes $ 0^{\circ} <b <2^{\circ} $ (for example, event IC14 \citep{2019ICRC...36..836A}) and at the periphery of the FBs are of particular interest. Monitoring of neutrino events was also carried out by ANTARES and KM3Net \citep{2017JPhCS.888a2102H}. We would like to conclude that high-energy neutrino analysis is a task for the near future. This will most accurately allow registering accelerating high-energy protons.
    
\section{Acknowledgments}
    
    Fruitful discussion with and comments by Dmitry Chernyshov are gratefully acknowledged. This work was supported by Russian Foundation for Fundamental Research, grant number 20-02-00469.
    
\section*{Data Availability}
    
    No new data were generated or analysed in support of this research.

\bibliographystyle{mnras}
\bibliography{References}

\begin{thebibliography}{}
\makeatletter
\relax
\def\mn@urlcharsother{\let\do\@makeother \do\$\do\&\do\#\do\^\do\_\do\%\do\~}
\def\mn@doi{\begingroup\mn@urlcharsother \@ifnextchar [ {\mn@doi@}
  {\mn@doi@[]}}
\def\mn@doi@[#1]#2{\def\@tempa{#1}\ifx\@tempa\@empty \href
  {http://dx.doi.org/#2} {doi:#2}\else \href {http://dx.doi.org/#2} {#1}\fi
  \endgroup}
\def\mn@eprint#1#2{\mn@eprint@#1:#2::\@nil}
\def\mn@eprint@arXiv#1{\href {http://arxiv.org/abs/#1} {{\tt arXiv:#1}}}
\def\mn@eprint@dblp#1{\href {http://dblp.uni-trier.de/rec/bibtex/#1.xml}
  {dblp:#1}}
\def\mn@eprint@#1:#2:#3:#4\@nil{\def\@tempa {#1}\def\@tempb {#2}\def\@tempc
  {#3}\ifx \@tempc \@empty \let \@tempc \@tempb \let \@tempb \@tempa \fi \ifx
  \@tempb \@empty \def\@tempb {arXiv}\fi \@ifundefined
  {mn@eprint@\@tempb}{\@tempb:\@tempc}{\expandafter \expandafter \csname
  mn@eprint@\@tempb\endcsname \expandafter{\@tempc}}}

\bibitem[\protect\citeauthoryear{{Ahlers}, {Bai}, {Barger}  \& {Lu}}{{Ahlers}
  et~al.}{2016}]{2016PhRvD..93a3009A}
{Ahlers} M.,  {Bai} Y.,  {Barger} V.,   {Lu} R.,  2016, \mn@doi [Phys.Rev.D]
  {10.1103/PhysRevD.93.013009}, \href
  {https://ui.adsabs.harvard.edu/abs/2016PhRvD..93a3009A} {93, 013009}

\bibitem[\protect\citeauthoryear{{Alvarez Hurtado}, {Fraija}, {Galv{\'a}n}  \&
  {Marinelli}}{{Alvarez Hurtado} et~al.}{2019}]{2019ICRC...36..836A}
{Alvarez Hurtado} P.,  {Fraija} N.,  {Galv{\'a}n} A.,   {Marinelli} A.,  2019,
  in 36th International Cosmic Ray Conference (ICRC2019). p.~836 (\mn@eprint
  {arXiv} {1908.03613})

\bibitem[\protect\citeauthoryear{{Armillotta}, {Krumholz}, {Di Teodoro}  \&
  {McClure-Griffiths}}{{Armillotta} et~al.}{2019}]{2019MNRAS.490.4401A}
{Armillotta} L.,  {Krumholz} M.~R.,  {Di Teodoro} E.~M.,   {McClure-Griffiths}
  N.~M.,  2019, \mn@doi [MNRAS] {10.1093/mnras/stz2880}, \href
  {https://ui.adsabs.harvard.edu/abs/2019MNRAS.490.4401A} {490, 4401}

\bibitem[\protect\citeauthoryear{{Barkov} \& {Bosch-Ramon}}{{Barkov} \&
  {Bosch-Ramon}}{2014}]{2014A&A...565A..65B}
{Barkov} M.~V.,  {Bosch-Ramon} V.,  2014, \mn@doi [A\&A]
  {10.1051/0004-6361/201322743}, \href
  {https://ui.adsabs.harvard.edu/abs/2014A&A...565A..65B} {565, A65}

\bibitem[\protect\citeauthoryear{{Blandford} \& {Znajek}}{{Blandford} \&
  {Znajek}}{1977}]{1977MNRAS.179..433B}
{Blandford} R.~D.,  {Znajek} R.~L.,  1977, \mn@doi [MNRAS]
  {10.1093/mnras/179.3.433}, \href
  {https://ui.adsabs.harvard.edu/abs/1977MNRAS.179..433B} {179, 433}

\bibitem[\protect\citeauthoryear{{Bouyahiaoui}, {Kachelrie}  \&
  {Semikoz}}{{Bouyahiaoui} et~al.}{2020}]{2020PhRvD.101l3023B}
{Bouyahiaoui} M.,  {Kachelrie} M.,   {Semikoz} D.~V.,  2020, \mn@doi
  [Phys.Rev.D] {10.1103/PhysRevD.101.123023}, \href
  {https://ui.adsabs.harvard.edu/abs/2020PhRvD.101l3023B} {101, 123023}

\bibitem[\protect\citeauthoryear{{Carretti} et~al.,}{{Carretti}
  et~al.}{2013}]{2013Natur.493...66C}
{Carretti} E.,  et~al., 2013, \mn@doi [Nature] {10.1038/nature11734}, \href
  {https://ui.adsabs.harvard.edu/abs/2013Natur.493...66C} {493, 66}

\bibitem[\protect\citeauthoryear{{Cheng}, {Chernyshov}, {Dogiel}, {Ko}  \&
  {Ip}}{{Cheng} et~al.}{2011}]{2011ApJ...731L..17C}
{Cheng} K.~S.,  {Chernyshov} D.~O.,  {Dogiel} V.~A.,  {Ko} C.~M.,   {Ip} W.~H.,
   2011, \mn@doi [ApJ] {10.1088/2041-8205/731/1/L17}, \href
  {https://ui.adsabs.harvard.edu/abs/2011ApJ...731L..17C} {731, L17}

\bibitem[\protect\citeauthoryear{{Cheng}, {Chernyshov}, {Dogiel}, {Ko}, {Ip}
  \& {Wang}}{{Cheng} et~al.}{2012}]{2012ApJ...746..116C}
{Cheng} K.~S.,  {Chernyshov} D.~O.,  {Dogiel} V.~A.,  {Ko} C.~M.,  {Ip} W.~H.,
   {Wang} Y.,  2012, \mn@doi [ApJ] {10.1088/0004-637X/746/2/116}, \href
  {https://ui.adsabs.harvard.edu/abs/2012ApJ...746..116C} {746, 116}

\bibitem[\protect\citeauthoryear{{Cheng}, {Chernyshov}, {Dogiel}  \&
  {Ko}}{{Cheng} et~al.}{2014}]{2014ApJ...790...23C}
{Cheng} K.~S.,  {Chernyshov} D.~O.,  {Dogiel} V.~A.,   {Ko} C.~M.,  2014,
  \mn@doi [ApJ] {10.1088/0004-637X/790/1/23}, \href
  {https://ui.adsabs.harvard.edu/abs/2014ApJ...790...23C} {790, 23}

\bibitem[\protect\citeauthoryear{{Cheng}, {Chernyshov}, {Dogiel}  \&
  {Ko}}{{Cheng} et~al.}{2015a}]{2015ApJ...799..112C}
{Cheng} K.~S.,  {Chernyshov} D.~O.,  {Dogiel} V.~A.,   {Ko} C.~M.,  2015a,
  \mn@doi [ApJ] {10.1088/0004-637X/799/1/112}, \href
  {https://ui.adsabs.harvard.edu/abs/2015ApJ...799..112C} {799, 112}

\bibitem[\protect\citeauthoryear{{Cheng}, {Chernyshov}, {Dogiel}  \&
  {Ko}}{{Cheng} et~al.}{2015b}]{2015ApJ...804..135C}
{Cheng} K.~S.,  {Chernyshov} D.~O.,  {Dogiel} V.~A.,   {Ko} C.~M.,  2015b,
  \mn@doi [ApJ] {10.1088/0004-637X/804/2/135}, \href
  {https://ui.adsabs.harvard.edu/abs/2015ApJ...804..135C} {804, 135}

\bibitem[\protect\citeauthoryear{{Chernyshov}, {Cheng}, {Dogiel}  \&
  {Ko}}{{Chernyshov} et~al.}{2014}]{2014NuPhS.256..179C}
{Chernyshov} D.~O.,  {Cheng} K.~S.,  {Dogiel} V.~A.,   {Ko} C.~M.,  2014,
  \mn@doi [Nuclear Physics B Proceedings Supplements]
  {10.1016/j.nuclphysbps.2014.10.021}, \href
  {https://ui.adsabs.harvard.edu/abs/2014NuPhS.256..179C} {256, 179}

\bibitem[\protect\citeauthoryear{{Chernyshov}, {Cheng}, {Dogiel}  \&
  {Ko}}{{Chernyshov} et~al.}{2017}]{2017EPJWC.14504004C}
{Chernyshov} D.,  {Cheng} K.-S.,  {Dogiel} V.,   {Ko} C.-M.,  2017, in European
  Physical Journal Web of Conferences. p. 04004,
  \mn@doi{10.1051/epjconf/201714504004}

\bibitem[\protect\citeauthoryear{{Crocker}}{{Crocker}}{2012}]{2012MNRAS.423.3512C}
{Crocker} R.~M.,  2012, \mn@doi [MNRAS] {10.1111/j.1365-2966.2012.21149.x},
  \href {https://ui.adsabs.harvard.edu/abs/2012MNRAS.423.3512C} {423, 3512}

\bibitem[\protect\citeauthoryear{{Crocker} \& {Aharonian}}{{Crocker} \&
  {Aharonian}}{2011}]{2011PhRvL.106j1102C}
{Crocker} R.~M.,  {Aharonian} F.,  2011, \mn@doi [PhRvL]
  {10.1103/PhysRevLett.106.101102}, \href
  {https://ui.adsabs.harvard.edu/abs/2011PhRvL.106j1102C} {106, 101102}

\bibitem[\protect\citeauthoryear{{Fragione} \& {Loeb}}{{Fragione} \&
  {Loeb}}{2020}]{2020ApJ...901L..32F}
{Fragione} G.,  {Loeb} A.,  2020, \mn@doi [ApJ] {10.3847/2041-8213/abb9b4},
  \href {https://ui.adsabs.harvard.edu/abs/2020ApJ...901L..32F} {901, L32}

\bibitem[\protect\citeauthoryear{{Guo}, {Mathews}, {Dobler}  \& {Oh}}{{Guo}
  et~al.}{2012}]{2012ApJ...756..182G}
{Guo} F.,  {Mathews} W.~G.,  {Dobler} G.,   {Oh} S.~P.,  2012, \mn@doi [ApJ]
  {10.1088/0004-637X/756/2/182}, \href
  {https://ui.adsabs.harvard.edu/abs/2012ApJ...756..182G} {756, 182}

\bibitem[\protect\citeauthoryear{{HESS Collaboration}, {Abramowski},
  {Aharonian}  \& al.}{{HESS Collaboration} et~al.}{2016}]{2016Natur.531..476H}
{HESS Collaboration} {Abramowski} A.,  {Aharonian} F.,   al. 2016, \mn@doi
  [Nature] {10.1038/nature17147}, \href
  {https://ui.adsabs.harvard.edu/abs/2016Natur.531..476H} {531, 476}

\bibitem[\protect\citeauthoryear{{Hallmann}, {Eberl}  \& {KM3NeT
  Collaboration}}{{Hallmann} et~al.}{2017}]{2017JPhCS.888a2102H}
{Hallmann} S.,  {Eberl} T.,   {KM3NeT Collaboration} 2017, in Journal of
  Physics Conference Series. p. 012102, \mn@doi{10.1088/1742-6596/888/1/012102}

\bibitem[\protect\citeauthoryear{{Istomin}}{{Istomin}}{2014}]{2014NewA...27...13I}
{Istomin} Y.~N.,  2014, \mn@doi [NewAstr] {10.1016/j.newast.2013.08.001}, \href
  {https://ui.adsabs.harvard.edu/abs/2014NewA...27...13I} {27, 13}

\bibitem[\protect\citeauthoryear{Istomin \& Gunya}{Istomin \&
  Gunya}{2020a}]{PhysRevD.102.043010}
Istomin Y.~N.,  Gunya A.~A.,  2020a, \mn@doi [Phys. Rev. D]
  {10.1103/PhysRevD.102.043010}, 102, 043010

\bibitem[\protect\citeauthoryear{{Istomin} \& {Gunya}}{{Istomin} \&
  {Gunya}}{2020b}]{2020MNRAS.492.4884I}
{Istomin} Y.~N.,  {Gunya} A.~A.,  2020b, \mn@doi [MNRAS]
  {10.1093/mnras/staa104}, \href
  {https://ui.adsabs.harvard.edu/abs/2020MNRAS.492.4884I} {492, 4884}

\bibitem[\protect\citeauthoryear{{Ko}, {Breitschwerdt}, {Chernyshov}, {Cheng},
  {Dai}  \& {Dogiel}}{{Ko} et~al.}{2020}]{2020ApJ...904...46K}
{Ko} C.~M.,  {Breitschwerdt} D.,  {Chernyshov} D.~O.,  {Cheng} H.,  {Dai} L.,
  {Dogiel} V.~A.,  2020, \mn@doi [ApJ] {10.3847/1538-4357/abbda4}, \href
  {https://ui.adsabs.harvard.edu/abs/2020ApJ...904...46K} {904, 46}

\bibitem[\protect\citeauthoryear{{Lacki}}{{Lacki}}{2014}]{2014MNRAS.444L..39L}
{Lacki} B.~C.,  2014, \mn@doi [MNRAS] {10.1093/mnrasl/slu107}, \href
  {https://ui.adsabs.harvard.edu/abs/2014MNRAS.444L..39L} {444, L39}

\bibitem[\protect\citeauthoryear{{Lunardini}, {Razzaque}, {Theodoseau}  \&
  {Yang}}{{Lunardini} et~al.}{2014}]{2014PhRvD..90b3016L}
{Lunardini} C.,  {Razzaque} S.,  {Theodoseau} K.~T.,   {Yang} L.,  2014,
  \mn@doi [PhRevD] {10.1103/PhysRevD.90.023016}, \href
  {https://ui.adsabs.harvard.edu/abs/2014PhRvD..90b3016L} {90, 023016}

\bibitem[\protect\citeauthoryear{{Mertsch} \& {Petrosian}}{{Mertsch} \&
  {Petrosian}}{2019}]{2019A&A...622A.203M}
{Mertsch} P.,  {Petrosian} V.,  2019, \mn@doi [A\&A]
  {10.1051/0004-6361/201833999}, \href
  {https://ui.adsabs.harvard.edu/abs/2019A&A...622A.203M} {622, A203}

\bibitem[\protect\citeauthoryear{{Razzaque} \& {Yang}}{{Razzaque} \&
  {Yang}}{2018}]{2018Galax...6...47R}
{Razzaque} S.,  {Yang} L.,  2018, \mn@doi [Galaxies] {10.3390/galaxies6020047},
  \href {https://ui.adsabs.harvard.edu/abs/2018Galax...6...47R} {6, 47}

\bibitem[\protect\citeauthoryear{{Su} \& {Finkbeiner}}{{Su} \&
  {Finkbeiner}}{2012}]{2012ApJ...753...61S}
{Su} M.,  {Finkbeiner} D.~P.,  2012, \mn@doi [ApJ]
  {10.1088/0004-637X/753/1/61}, \href
  {https://ui.adsabs.harvard.edu/abs/2012ApJ...753...61S} {753, 61}

\bibitem[\protect\citeauthoryear{{Su}, {Slatyer}  \& {Finkbeiner}}{{Su}
  et~al.}{2010}]{2010ApJ...724.1044S}
{Su} M.,  {Slatyer} T.~R.,   {Finkbeiner} D.~P.,  2010, \mn@doi [ApJ]
  {10.1088/0004-637X/724/2/1044}, \href
  {https://ui.adsabs.harvard.edu/abs/2010ApJ...724.1044S} {724, 1044}

\bibitem[\protect\citeauthoryear{{Yang}, {Ruszkowski}, {Ricker}, {Zweibel}  \&
  {Lee}}{{Yang} et~al.}{2012}]{2012ApJ...761..185Y}
{Yang} H. Y.~K.,  {Ruszkowski} M.,  {Ricker} P.~M.,  {Zweibel} E.,   {Lee} D.,
  2012, \mn@doi [ApJ] {10.1088/0004-637X/761/2/185}, \href
  {https://ui.adsabs.harvard.edu/abs/2012ApJ...761..185Y} {761, 185}

\bibitem[\protect\citeauthoryear{{Zhang} \& {Guo}}{{Zhang} \&
  {Guo}}{2020}]{2020ApJ...894..117Z}
{Zhang} R.,  {Guo} F.,  2020, \mn@doi [ApJ] {10.3847/1538-4357/ab8bd0}, \href
  {https://ui.adsabs.harvard.edu/abs/2020ApJ...894..117Z} {894, 117}

\makeatother
\end{thebibliography}
	
\end{document}